\documentclass{llncs}
\usepackage{makeidx}  
\usepackage{graphicx}
\usepackage{amsmath}
\usepackage{amssymb}
\usepackage[colorinlistoftodos]{todonotes} 
\usepackage{stmaryrd}
\usepackage{tabularx}

\begin{document}
\frontmatter          
\mainmatter              
\title{Improved MR to CT synthesis for PET/MR attenuation correction using Imitation Learning}

\author{Kerstin Kl\"{a}ser$^{1,2}$ \and Thomas Varsavsky$^{1,2}$ \and Pawel Markiewicz$^{1,2}$ \and Tom Vercauteren$^2$ \and David Atkinson$^4$ \and Kris Thielemans$^3$ \and Brian Hutton$^3$ \and M Jorge Cardoso$^2$ \and S\'{e}bastien Ourselin$^2$ }
\institute{$^1$Dept. Medical Physics \& Biomedical Engineering, University College London, UK \\ $^2$School of Biomedical Engineering \& Imaging Sciences, King's College London, UK \\ $^3$Institute of Nuclear Medicine, University College London, UK \\ $^4$Centre for Medical Imaging, University College London,  UK}
\maketitle

%
%
\begin{abstract}
The ability to synthesise Computed Tomography images - commonly known as pseudo CT, or pCT - from MRI input data is commonly assessed using an intensity-wise similarity, such as an \linebreak $L_2-norm$ between the ground truth CT and the pCT. However, given that the ultimate purpose is often to use the pCT as an attenuation map ($\mu$-map) in Positron Emission Tomography Magnetic Resonance Imaging (PET/MRI), minimising the error between pCT and CT is not necessarily optimal. The main objective should be to predict a pCT that, when used as $\mu$-map, reconstructs a pseudo PET (pPET) which is as close as possible to the gold standard PET. To this end, we propose a novel multi-hypothesis deep learning framework that generates pCTs by minimising a combination of the pixel-wise error between pCT and CT and a proposed metric-loss that itself is represented by a convolutional neural network (CNN) and aims to minimise subsequent PET residuals. The model is trained on a database of 400 paired MR/CT/PET image slices. Quantitative results show that the network generates pCTs that seem less accurate when evaluating the Mean Absolute Error on the pCT (69.68HU) compared to a baseline CNN (66.25HU), but lead to significant improvement in the PET reconstruction - 115\textit{a.u.} compared to baseline 140\textit{a.u}.
\end{abstract}

\section{Introduction}
 The combination of Positron Emission Tomography (PET) and Magnetic Resonance Imaging (MRI) marked a significant event in the field of Nuclear Medicine, facilitating simultaneous structural and functional characterisation of soft tissue \cite{pichler2008positron}. In order to accurately reconstruct quantitative PET images, it is indispensable to correct for attenuation of the whole imaging object (part of the human body) including the hardware (patient bed and additional coils). However, this is particularly challenging in PET/MRI as there is no direct correlation between MR image intensities and attenuation coefficients in contrast to the case when a CT image is available. In hybrid imaging systems that combine PET with Computed Tomography (CT), the tissue density information is derived from the CT image as Hounsfield units (HU), which can bi-linear approximate the attenuation coefficients ({$\mu$}). While CT remains the clinically accepted gold-standard for PET/MR attenuation correction, it is desirable to generate accurate {$\mu$}-maps without the need for an additional CT acquisition. Hence, the concept of synthesising pseudo CT (pCT) images from MRs raised significant attention in the research area of PET/MR reconstruction.
 
 Recently, a multi-centre study has shown that compared to physics and segmentation based approaches, methods based on multi-atlas approaches were best suited to generate appropriate pCTs. These methods estimate {$\mu$}-maps on a continuous scale by deforming an anatomical model that contains paired MR and CT data to match the subject's anatomy by using non-rigid registration algorithms \cite{ladefoged2017multi}. 
 
\begin{figure}[t!]
    \centering
    \includegraphics[width=0.9\textwidth]{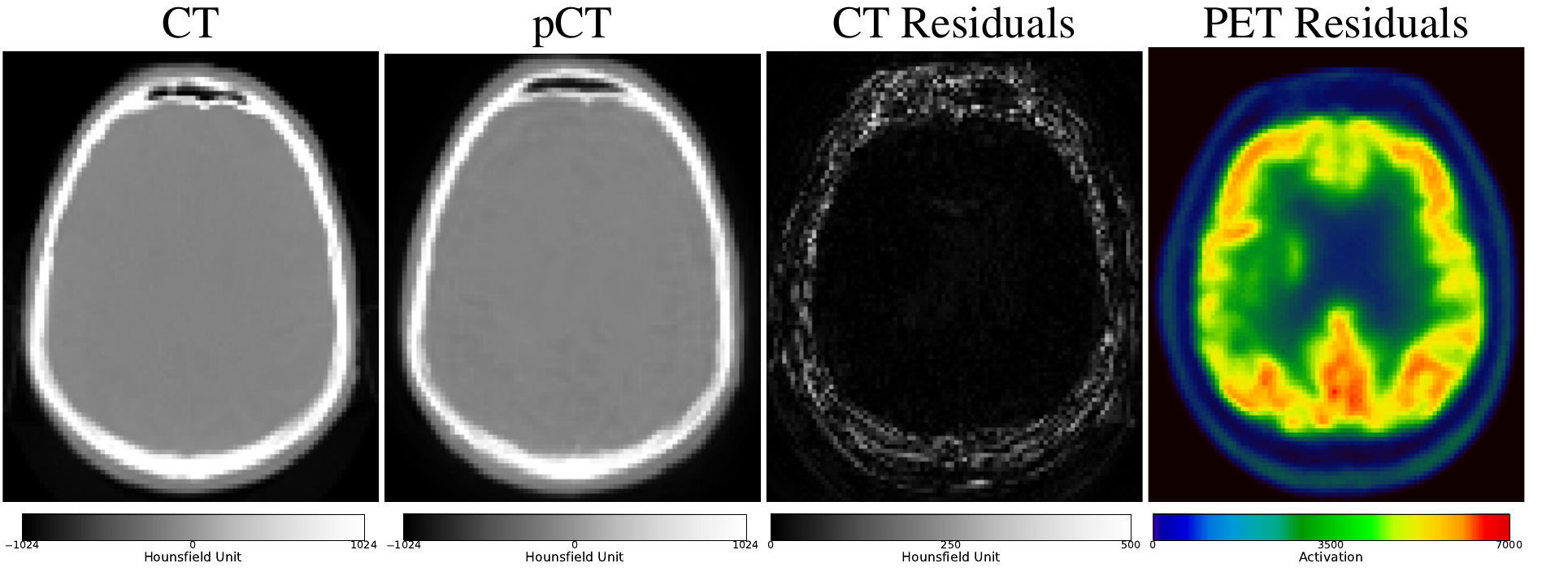}
     \caption{a) The ground truth CT, b) the predicted pseudo CT, c) the absolute residual between true and pseudo CT, and d) the absolute residual between PETs reconstructed using the CT and synthetic CT as attenuation maps. Note that small and very localised difference in the CT (c) result in large PET residuals (d). We argue that algorithms should be optimising for PET residuals (d) and not for CT residuals (c).} 
    \label{fig:error_diff}
\end{figure}

 In recent years, there has been a shift of emphasis in the field of PET/MR attenuation correction towards deep learning approaches that have demonstrated significant improvements in the MR to CT image translation task, surpassing state-of-the-art multi-atlas-based approaches \cite{Ninon}. Such methods often utilise convolutional neural networks (CNN) that are able to capture the contextual information between two image domains (as between MR and CT) in order to translate one possible representation of an image into another. Supervised learning settings assume that the training dataset comprises examples of an input image (e.g. MR here) along with their corresponding target image (i.e. CT here). A popular method to optimise image translation networks is to minimise the residuals between the predicted pCT and the corresponding ground-truth CT, equivalent to minimising the $\mathcal{L}_2$-loss. $\mathcal{L}_2$-losses make sense when the optimal pCT for PET reconstruction is the one that is the closest, intensity-wise, to the target ground truth CT. However, this $\mathcal{L}_2$-loss fails to recognise that the primary objective of CT synthesis is to create a synthetic CT that, when used to reconstruct the PET image, makes it as close as possible to the gold standard PET reconstructed with the true CT. Also, the risk-minimising nature of the $\mathcal{L}_2$-loss disregards the fact that small local differences between the pCT and the true CT can have a large impact on the reconstructed PET. This downstream impact in PET reconstruction is illustrated in Fig.\ref{fig:error_diff}.
 
 With the emergence of the cycleGAN in 2017 \cite{CycleGAN}, many efforts have been made to synthesise CT images in an unsupervised manner disregarding the need of the $\mathcal{L}_2$-loss. Wolterink et al. \cite{Wolternik} used a CNN that minimises an adversarial loss to learn the mapping from MR to CT. This adversarial loss forces the pseudo CT to be indistinguishable from a real CT. A second CNN ensures that the pseudo CT corresponds to the actual input MR image. However, using a cycleGAN only to synthesise pseudo CTs does not necessarily guarantee structural consistency between pseudo CT and original CT. Therefore Yang et al. \cite{Yang} proposed a structure-constraint cycleGAN that minimises an additional structural consistency loss. In 2019, Jin et al. \cite{Jin} presented a method that combines paired and unpaired data in order to overcome the missing structural consistency of the cycleGAN and to mitigate the errors introduced by the registration of paired data. 
 
 To the best of our knowledge, all these methods focus on minimising the error of the synthesised pCT. However, synthesising a CT image only acts as an interim step when aiming for PET attenuation correction creating an additional stage for potentially introduced errors. This work aims to directly minimise the PET residuals and achieves this by introducing a novel MR to CT synthesis framework that is composed of two separate CNNs. The first network generates multiple plausible CT representations using Multi-Hypothesis Learning instead of just a single pCT \cite{rupprecht2017learning}. An oracle determines the most correct predictor and only updates the weights with regards to the winning mode, enabling the first network to specialise in generating pCTs with specific features (e.g. skull thickness, bone density). A second network then predicts the residuals between ground-truth PETs and PETs reconstructed using each plausible pCT using imitation learning. In this setting, the second network can be seen as a metric that estimates the pPET residuals, and thus, by minimising this metric, the network learns to generate pCTs that will subsequently result in pPETs with lower residual errors.    
 
 \section{Methods}
 
 \subsection{Multi-Hypothesis Learning}
 
 Given a set of input MR images $x \in \mathcal{X}$ and a set of output CT images $y \in \mathcal{Y}$, the proposed image synthesis approach aims to find a mapping function $f_\phi$ between the two image domains $\mathcal{X}$ and $\mathcal{Y}$ , i.e. $f_\phi: \mathcal{X} \rightarrow \mathcal{Y} \quad \textrm{with unique parameters} \linebreak \phi \in \mathbb{R}^n$. In a supervised learning setting with a set of $N$ paired training tuples $(x_i , y_i)$, $i=1,...,N$, we try to find the predictor $f_\phi$ that minimises the error
 
 \[
 \frac{1}{N} \sum_{i=1}^{N} \mathcal{L}(f_\phi(x_i),y_i).
 \]
 
\noindent where $\mathcal{L}$ can be any desired loss, such as the classical $\mathcal{L}_2$-loss. In the proposed multi-hypothesis scenario, the network provides multiple predictions pCT, where $f_\phi^{j}(x) \in (f_{\phi}^{1}(x), ..., f_{\phi}^{M}(x)) \quad \textrm{with} \quad M \in \mathbb{N}$.
 
\noindent As in the original work \cite{rupprecht2017learning}, only the loss of the best predictor $f_{\phi}^{j}(x)$ will be considered in the training following a Winner-Takes-All (WTA) strategy, i.e. 
 
\[
\mathcal{L}(f_{\phi}(x_i), y_i) = min_{j\in [1,M]}\mathcal{L}(f^{j}_{\phi}(x_i),y_i) .
\]

\noindent This way the network learns $M$ modes to predict pCTs each specialising on specific features.
 
\subsection{Imitation Learning}
 
Following the hypothesis that the $\mathcal{L}_2$-loss is not an optimal loss metric when generating pCTs for the purpose of PET/MR attenuation correction because of its risk minimising nature, we propose to train a second network that, by taking ground truth CTs ($y_i$) and pCTs ($f_{\phi}^{j}(x_i) \in \tilde{\mathcal{Y}}$) as inputs, aims to approximate the function $g_\psi:\mathcal{Y},\tilde{\mathcal{Y}}\rightarrow \mathcal{Z} \quad \textrm{with} \quad \psi \in \mathbb{R}^n$. Here, $\mathcal{Z}$ is a set of error maps between the ground truth PET and the pPET that was reconstructed using each of the $M$ pCT realisations as $\mu$-maps. In other words, this second network tries to predict what the PET residuals would be from an input CT-pCT pair, thus imitating, or approximating, the PET reconstruction process. We train this imitation network by minimising the $\mathcal{L}_2$-loss between the true PET uptake error $z$ and the predicted error $\tilde{z}$, i.e. $ \mathcal{L}_2 = || z - \tilde{z} ||_2$.
 
 Lastly, we use this second network as a new loss function for the first network, as it provides an approximate and differentiable estimate of the PET residual loss. The loss minimised by the first network is then defined as \linebreak $\mathcal{L}(x_i, y_i, z_i) = \min_{m \in [1,M]} [g_\psi(f_\phi(x_i),y_i),z_i]$.
 
 \subsection{Proposed Network Architecture}
 
  \begin{figure}[b!]
    \centering
    \includegraphics[width=\textwidth]{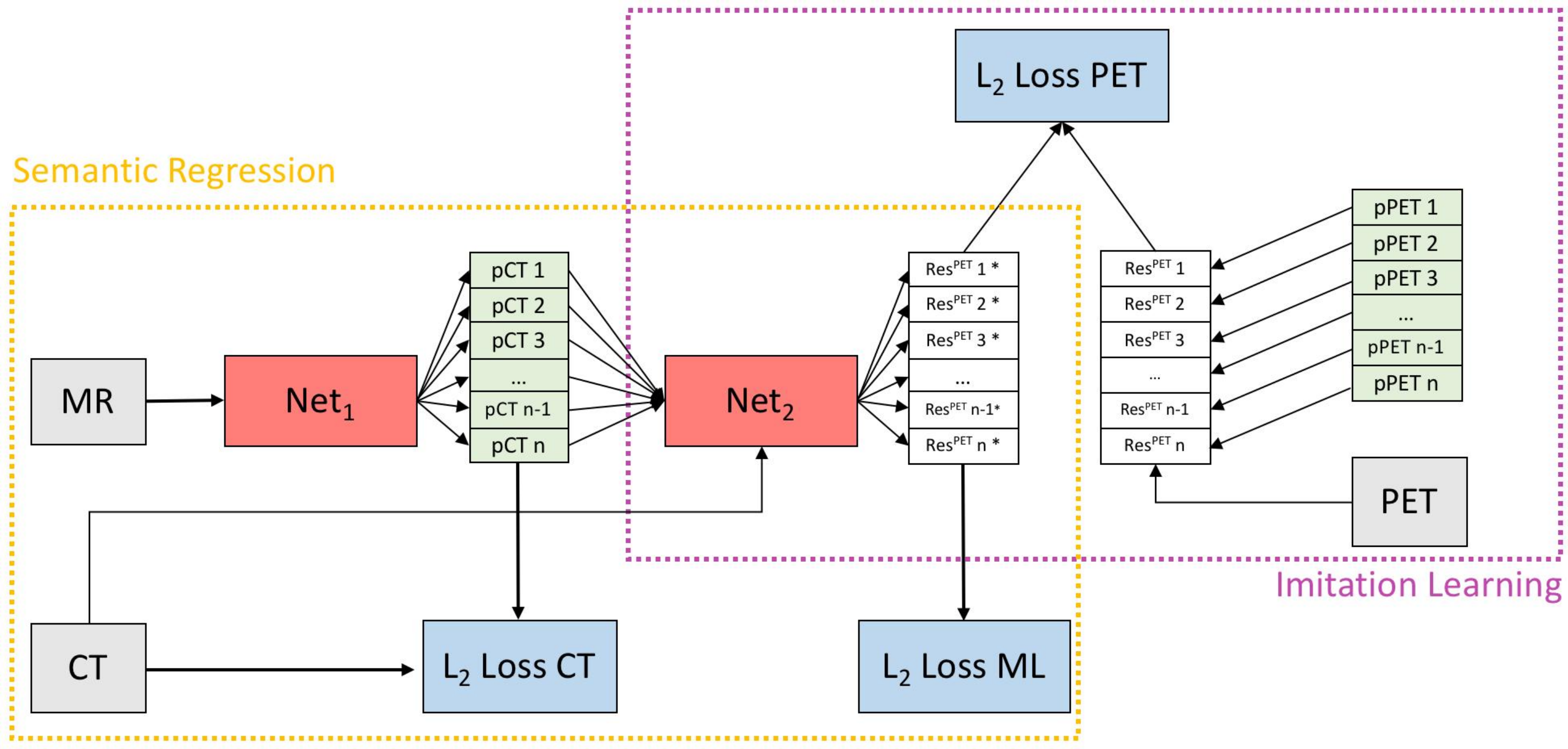}
    \caption{Yellow box: semantic regression. $Net_1$ takes MR images as inputs and predicts multiple pCT realisations by minimising a combination of the $\mathcal{L}_2$-loss between ground truth CT and pCT ($\mathcal{L}_2$-loss CT) and a learned metric loss ($\mathcal{L}_2$-loss ML). In the first stage only $\mathcal{L}_2$-loss CT is considered and $\mathcal{L}_2$-loss ML is weighted to zero. Purple box: imitation network. $Net_2$ takes pCTs and corresponding CTs as input and predicts the error between ground truth PET and pPET reconstructed with pCT as $\mu$-map by minimising $\mathcal{L}_2$-loss PET. Semantic regression and imitation network are trained separately in three stages.} 
    \label{fig:architecture}
\end{figure}

The proposed network architecture (Fig. \ref{fig:architecture}) is trained in three phases: First, a HighResNet \cite{Wenqi} with multiple hypothesis outputs is trained with $\mathcal{L}_2$-WTA loss to generate different pCT (yellow box). In the second stage, while freezing the weights of the first network a second instance (purple box) of HighResNet is trained to learn the error prediction between true and predicted PET and learn the mapping between pCT residual and subsequent pPET reconstruction error. Once learnt, the first network is retrained using both the CT $\mathcal{L}_2$-loss and the metric loss in equal proportions.

 
 \subsection{Implementation Details}
 The training was performed on whole images using 70\% of the dataset (10\% was reserved for validation and 20\% for testing). All training phases were performed on a Titan V GPU with Adam optimiser. A model was trained with a learning rate of 0.001 for 20k iterations decreasing the learning rate by a factor of 10 and resuming training until convergence. The architecture was implemented using NiftyNet, an open-source TensorFlow-based CNN platform developed for research in the domain of medical image analysis \cite{NiftyNet}.

\section{Experimental Datasets and Materials}
\label{section3}
The experimental dataset consisted of pairs of T1- and T2-weighted MR and CT brain images of 20 patients. For each subject, an intra-subject registration was performed, where MRs and CTs were aligned using first a rigid registration algorithm followed by a very low degree of freedom non-rigid deformation \cite{Ninon}. A second non-linear registration was performed, using a cubic B-spline with normalised mutual information, only on the neck region to correct for soft tissue shift \cite{Modat}. Each volume had 301 x 301 x 153 voxels with a voxel size of approximately 1$mm^3$. For the purpose of this work, the original data was then resampled to the original Siemens Biograph mMR PET resolution of 344 x 344 x 127 voxels with a voxel size of approximately 2$mm^3$ before we extracted the 20 central slices per volume resulting in a registered 2D MR/CT/PET dataset of 400 images. MR and CT images were downsampled because all image analysis was performed in the original PET space since the ultimate aim of the method is to minimise PET residuals. For evaluation purposes, a head region mask was extracted from the CT image to exclude the background from the performance metric analysis. In order to train the imitation network, three PETs were reconstructed using each of the multi-hypothesis pCTs over 20 slices (here denoted as pPET), resulting in a total of 60 pCT/pPET pairs. PET reconstruction was performed using NiftyPET \cite{Pawel}. Since the raw PET data was not accessible, the following simulation was performed: a PET forward projection was applied on the $\mu$-map transformed versions of the pCTs in order to obtain attenuation factor sinograms. Similar forward projection was applied to the original PET images to obtain simulated emission sinograms which are then attenuated through element-wise multiplication using the attenuation factor sinograms. Those simulated sinograms were then reconstructed using both the original CT-based attenuation map to obtain a reference image, as well as the attenuation maps derived from the different pCT images.

\section{Experiments and Results}

\begin{figure}[h!]
    \centering
    \includegraphics[width=0.9\textwidth]{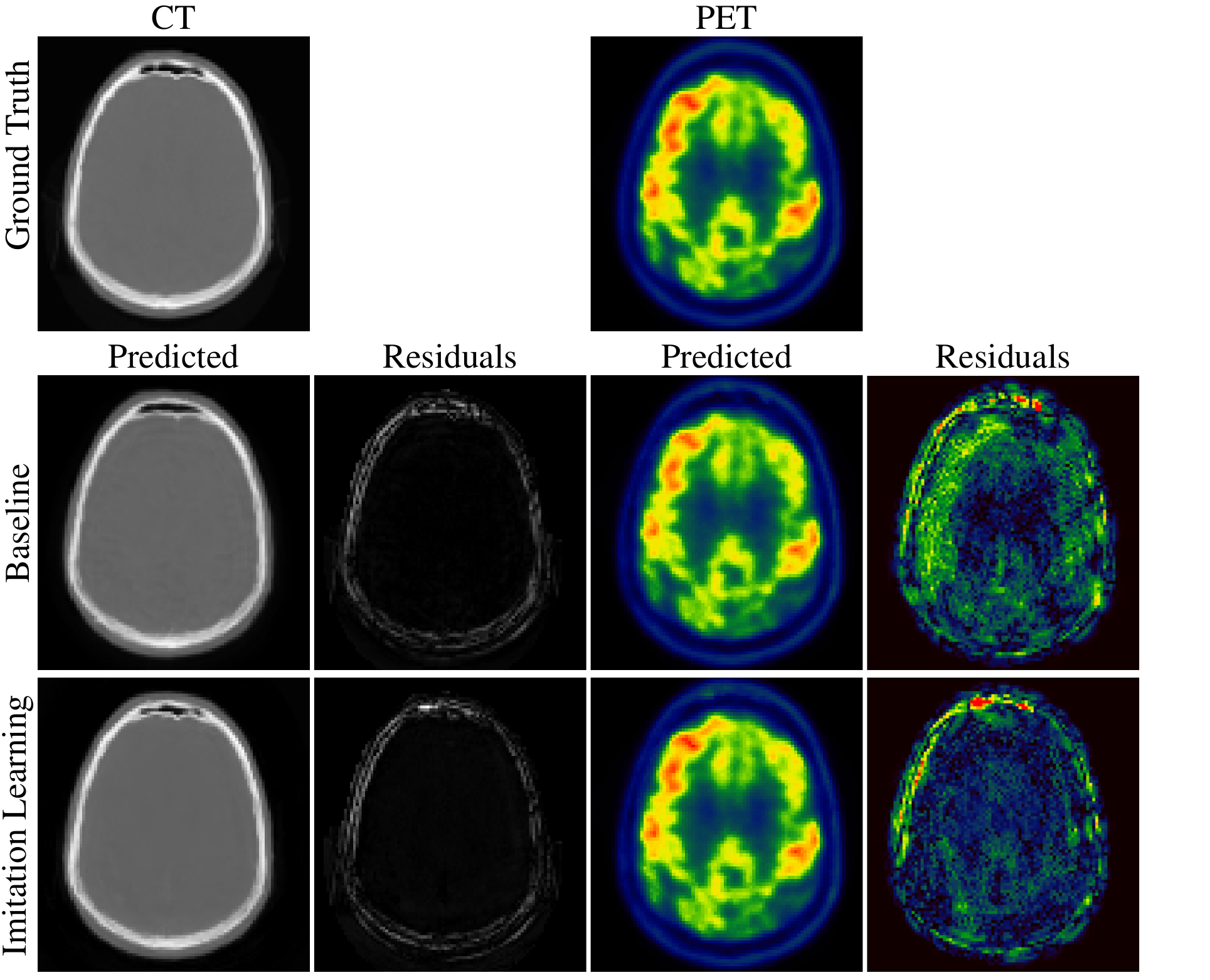}
    \caption{Qualitative results. From top to bottom: Ground-truth, baseline (HighResNet), and imitation learning. From left to right: CT, pCT-CT residuals, PET, pPET-PET residuals. As expected, we note that MAE in the pCT generated with the proposed imitation learning is higher than the baseline, but the resulting pPET error is significantly \textit{lower} for the proposed method.} 
    \label{fig:qualitative_results}
\end{figure}

Qualitative results are presented in Fig. \ref{fig:qualitative_results}. The first column shows the ground truth CT image (top), the pCTs generated with the HighResNet that we used as baseline (middle) and a pCT generated with the proposed imitation learning (bottom). Next to the CTs (2nd column) the error between pCT and ground truth CT is shown. In the third column the true PET (top), imitation learning pPET (middle) and the baseline pPET (bottom) are shown followed by the corresponding pPET residuals in the last column. 

\begin{figure}[t!]
    \centering
    \includegraphics[width=0.8\textwidth]{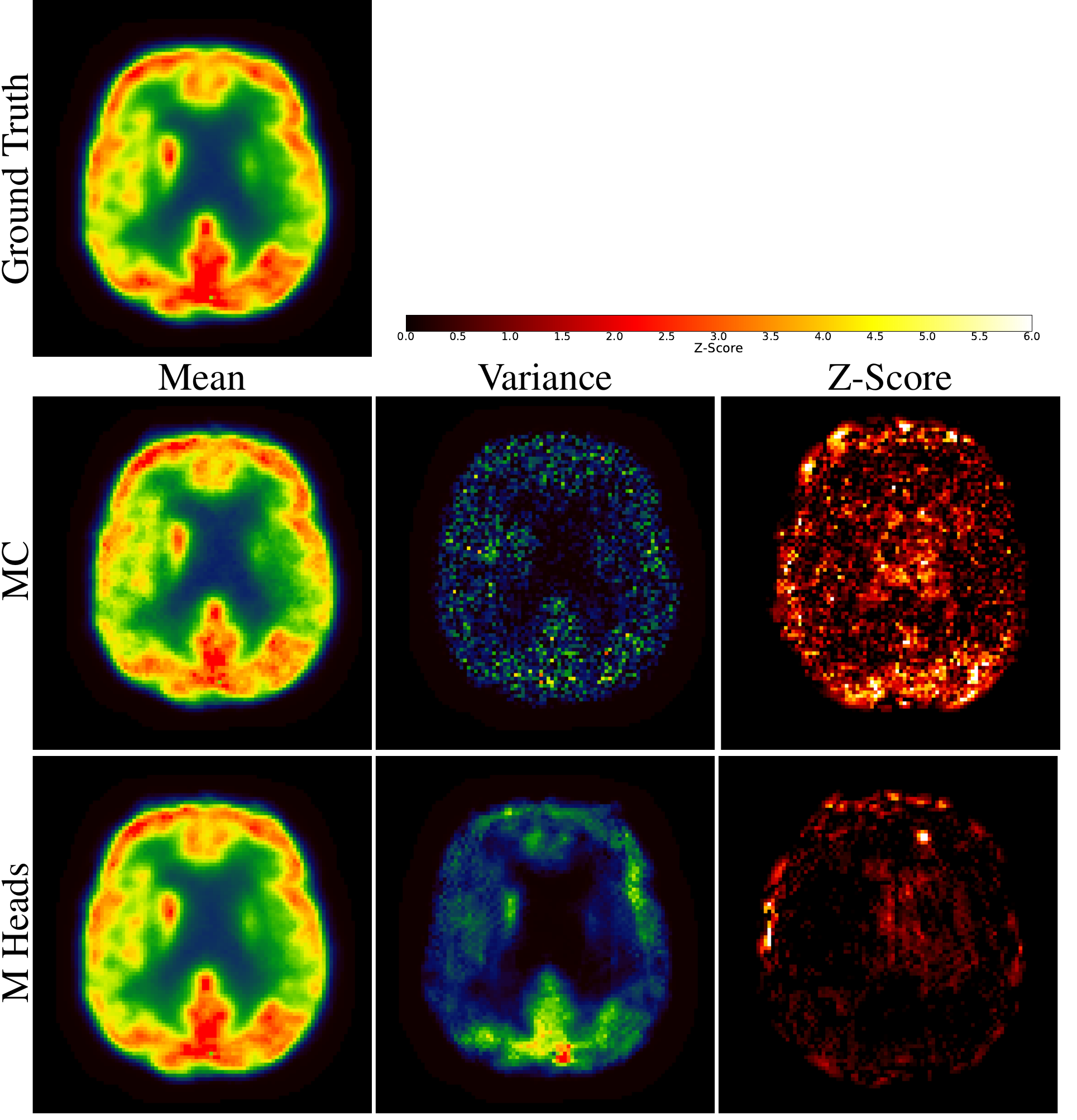}
    \caption{PET intensities (first column), variance (middle column) and Z-score (right column) of ground truth PET (top row) compared to pPET intensities reconstructed with pCTs from Monte Carlo (MC) dropout sampling (middle row) and pCTs from multi-hypothesis sampling (bottom row). Sampling from multi-hypothesis captures true PET intensities better than sampling from MC dropout.} 
    \label{fig:MCvsMHeads}
\end{figure}

As a second experiment, we performed an evaluation on the use of Monte-Carlo (MC) dropout versus multi-hypothesis as a sampling scheme to generate multiple realisations of pCTs. The results are depicted in Fig. \ref{fig:MCvsMHeads}. The variance in the pPET intensities, which was reconstructed with a $\mu$-map from the pCTs generated with MC dropout, was found to be artificially low, while the multiple pCT realisations of the proposed multi-hypothesis model provided a wider distribution of pPET intensities. In order to investigate the accuracy of the predictions, we investigated the Z-score of both sampling schemes in order to show the relationship of the mean data distribution to the ground truth PET. \linebreak Fig. \ref{fig:MCvsMHeads}-Right presents the per pixel Z-score defined as $\tfrac{\text{PET}-\mu(\text{pPET}^M)}{\sigma(\text{pPET}^M)}$, with \linebreak $\mu(\text{pPET}^M)$ and $\sigma(\text{pPET}^M)$ being the per-pixel average and per pixel variance over $M$ pPET samples respectively. Results show that the Z-score for multi-hypothesis is significantly lower in the brain region than the one from MC dropout, meaning that the multi-hypothesis-based PET uncertainty does encompass the true PET value more often than the competing MC dropout method.  

In a third experiment, and for quantification purposes, we calculated the Mean Absolute Error (MAE) of the pseudo CTs only in the head region and of the pseudo PET only in the brain region by masking out the background of the images. We validated the advantages of the proposed imitation learning model on the remaining 20\% of the dataset hold out for testing (see Table \ref{MAE}). Although, as expected, the proposed method leads to a higher MAE on the CT (69.68 $\pm$ 32.22HU) compared to the simple feed forward model (66.25 $\pm$ 30.54HU), the MAE in the resulting pPET is significantly lower (paired t-test, $p<10^{-4}$) for the proposed method (115.41 $\pm$ 78.72) when compared to the baseline model (140.76 $\pm$ 91.87). The proposed method also outperforms the multi-hypothesis only approach in both metrics. 

\begin{table}[h!]   
\begin{center}
	\renewcommand{\arraystretch}{1.3}
\begin{tabularx}{\textwidth}{|@{}l|X|X|X@{}}
    \hline
    Method & MAE CT (in HU) & MAE PET (in a.u.)\\ 
    \hline 
    HighResNet & 66.25 $\pm$ 30.54 & 140.76 $\pm$ 91.87  \\
    Multi-Hypothesis  & 72.23 $\pm$ 27.69 & 215.57 $\pm$ 102.99 \\
    Imitation Learning & 69.68 $\pm$ 32.22 & 115.41 $\pm$ 78.72 \\
    \hline
    \end{tabularx}
    \caption{Mean Absolute Error (MAE) pCTs generated with HighResNet, Multi-Hypothesis pCTs and Imitation Learning pCTs and corresponding MAE in pPET.}
    \label{MAE}
\end{center}
\end{table}

\section{Discussion and Conclusion}
In this work, we proposed a novel network architecture for pCT synthesis for PET/MR attenuation correction. We were able to show that the $\mathcal{L}_2$-loss, often used as a minimisation metric in the field of CT synthesis, is not optimal when ultimately aiming for a low error in the corresponding pPET when used as attenuation map. Quantitative analysis on an independent dataset confirmed the proposed hypothesis that pCTs with a low MAE do not necessarily result in a low pPET error. This work also demonstrates that minimising a more suitable metric that indeed optimises for PET residuals (from CTs and pCTs) can improve the process of CT synthesis for PET/MR attenuation correction. 

\section*{Acknowledgements}
This work was supported by an IMPACT studentship funded jointly by \linebreak Siemens and the EPSRC UCL CDT in Medical Imaging (EP/L016478/1). We gratefully acknowledge the support of NVIDIA Corporation with the donation of one Titan V. This project has received funding from Wellcome Flagship Programme (WT213038/Z/18/Z), the Wellcome EPSRC CME (WT203148/Z/16/Z), the NIHR GSTT Biomedical Research Centre, and the NIHR UCLH Biomedical Research Centre.


\bibliographystyle{splncs}
\bibliography{ReferencesTech}

\end{document}